\documentclass{elsart}
\usepackage{natbib}

\usepackage{epsfig}

\usepackage{amssymb}

\begin{document}

\begin{frontmatter}

\title{Near-infrared integral-field spectroscopy of violent starburst
environments}

\author{Richard de Grijs}

\address{Department of Physics \& Astronomy, The University of
Sheffield, Hicks Building, Hounsfield Road, Sheffield S3 7RH, UK}

\begin{abstract}
Near-infrared (NIR) integral-field spectroscopy (IFS) of violent
starburst environments at high spatial (and spectral) resolution has
the potential to revolutionise our ideas regarding the local
interactions between the newly-formed massive stars and the
interstellar medium (ISM) of their host galaxies. To illustrate this
point, I present NIR IFS analysis of the central starburst region of
NGC 1140, obtained with CIRPASS on Gemini-South. While strong [Fe{\sc
ii}] emission is found throughout the galaxy, higher-order Brackett
emission is predominantly associated with the northern starburst
region. Based on the spatial distributions of the [Fe{\sc ii}] versus
Brackett line emission, I conclude that a galaxy-wide starburst was
induced several $\times 10^7$ yr ago, with more recent starburst
activity concentrated around the northern starburst region. I look
forward and discuss the exciting prospects that IFS at higher spatial
(and spectral) resolution will allow us trace (i) the massive outflows
(``superwinds'') expected to originate in the dense, young massive
star clusters commonly found in intense starburst environments, and
(ii) their impact on the galaxy's ISM.
\end{abstract}


\end{frontmatter}

\section{NGC 1140: A case for galactic cannibalism?}

\begin{figure}
\begin{center}
\psfig{figure=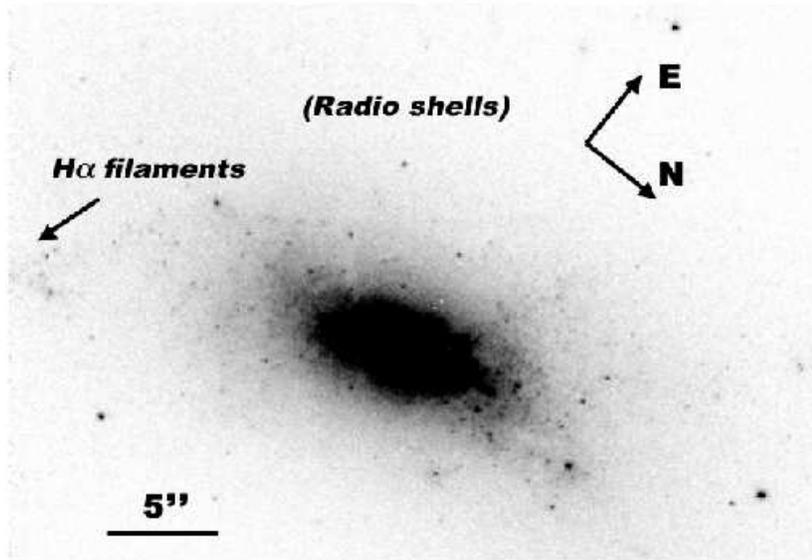, width=13cm}
\end{center}
\caption{\label{n1140image.fig}NGC 1140 as seen with the {\sl Hubble
Space Telescope} through the F555W ($V$-band) filter. The contrast has
been chosen to show the low-level structures; the location of the
radio shells identfied by Hunter et al. (H94a) is indicated, as is the
direction in which H$\alpha$ filaments extend (e.g., Jansen 2004).}
\end{figure}

The southern starburst dwarf galaxy NGC 1140 is remarkable in the
number of compact, young, massive star clusters (YMCs) that it has
formed recently in the small volume of its bright emission-line
core. Based on H{\sc i} and optical observations [e.g., Hunter,
O'Connell \& Gallagher 1994a [hereafter H94a], de Grijs et al. 2004],
and comparison with models (Hunter, van Woerden \& Gallagher 1994b,
hereafter H94b), it exhibits the characteristics of a recent merger
event, which has presumably induced the YMC formation in its centre,
and which is currently in its final stages. The galaxy is unusually
bright for its dwarf (or amorphous) morphological type, and at faint
H{\sc i} and optical light levels it shows multiple, misaligned
shell-like structures (H94a; see also Fig. \ref{n1140image.fig}).
These are reminiscent of the shells associated with some elliptical
galaxies that are thought to be remnants of a past tidal encounter
(e.g., Schweizer \& Seitzer 1988, Hernquist \& Spergel 1992).

The galaxy is also unusual in its content and overall morphology: its
optical appearance is dominated by a supergiant H{\sc ii} region
encompassing most of its centre, with an H$\alpha$ luminosity far
exceeding that of the giant H{\sc ii} region 30 Doradus in the Large
Magellanic Cloud (H94b). Its gas fraction, gas-to-luminosity ratio,
and H{\sc i} velocity dispersion are unusually high, and its H$\alpha$
velocity profiles remarkably broad, by roughly an order of magnitude,
for the mid-spiral type galaxy one would expect, based on its total
mass (H94b).

In addition, the formation of compact YMCs seems to be a hallmark of
intense episodes of recent, active star formation. Compact YMCs are
therefore important as probes of their host galaxy's recent
star-formation history, and also of its chemical evolution, the
stellar initial mass function (IMF), and other physical
characteristics in starbursts.

Although NGC 1140 has a small dwarf companion galaxy (which might
suggest that gravitational interactions with this companion may have
induced part of the recent active star formation), these observations
led Hunter et al. (H94b) to conclude that the galaxy is most likely
the product of the accretion of a low surface brightness gas-rich
companion by a relatively normal mid-type spiral galaxy of roughly
equal mass in the last $\sim 1$ Gyr. This picture is consistent with
the low metallicities and complex H{\sc i} morphology (including a
warped disc) and velocity field.

\section{The promise of integral-field spectroscopy}

NGC 1140 is in many ways similar to the prototype starburst galaxy
M82, in which we discovered a large population of $\sim 1$ Gyr-old
YMCs (de Grijs et al. 2001, 2003). It does not, however, feature a
minor-axis superwind as in M82.

The interplay between the YMCs and the galaxy's interstellar medium
(ISM) is, as in most active starburst galaxies, highly complex. Unlike
in M82, however, the average internal extinction appears to be
moderate (see de Grijs et al. 2004 for an overview). Nevertheless, in
order to (i) minimise the effects of residual internal extinction, and
(ii) sample the highly complex starburst environment at the highest
spatial {\it and} spectral resolution, we obtained near-infrared
imaging spectroscopy with the Cambridge InfraRed Panoramic Survey
Spectrograph (CIRPASS; Parry et al. 2000), which was deployed at
Gemini-South as part of the instrument commissioning/demonstration
science programme, carried out in Director's Discretionary Time (on UT
6 August 2002).

We aimed at obtaining significantly more accurate age estimates for
the YMCs than broad-band {\sl Hubble Space Telescope} observations
would allow us (cf. H94a, de Grijs et al. 2004), particularly for the
more obscured YMCs. This would allow us to independently verify their
tidal interaction-induced starburst origin (e.g. H94b). In addition,
the two-dimensional spectroscopic sampling would -- for the first time
-- allow us to assess the state and age of the disk stellar population
and the ISM. Thus, we expected that our CIRPASS data would allow us to
construct an ``age map'' of NGC 1140's central starburst region, which
could then be compared with the dynamical models for the
interaction-induced starburst (e.g. H94b).

CIRPASS is a NIR fibre-fed spectrograph, connected to a 490-element
integral-field unit (IFU). The variable lenslet scale was set to 0.36
arcsec diameter, and the hexagonal lenslets are arranged in the IFU to
survey an approximately rectangular area of $13.0 \times 4.7$ arcsec,
which provides sufficient sampling of the unusually bright diffuse
light in NGC 1140. A 400 {\it l} mm$^{-1}$ grating was used, which
produced a dispersion of 2.25{\AA} pixel$^{-1}$ and a wavelength
coverage of $1.45 - 1.67 \mu$m, with a resolving power of $R \equiv
(\lambda / \Delta\lambda)_{\rm FWHM} = 3500$.

The observations were taken in non-photometric conditions, with a
seeing of $\approx 1.0-1.5$ arcsec FWHM. We obtained three 40-minute
exposures of the galaxy, and one 40-minute offset sky exposure.
Although we had noticed that our last 40-minute exposure was affected
by high winds (and therefore blurred), during the analysis stage we
found that the guiding of one of the other exposures had experienced
problems as well due to wind shake. Thus, the CIRPASS results
discussed here are based on the single 40-minute exposure obtained
under the best available conditions.

\section{Propagation of star-formation activity?}

The wavelength range from $1.45 - 1.67 \mu$m covers a number of
emission lines expected to be strong in active starburst regions. We
identified four strong emission lines in a significant fraction of the
lenslets covering the NGC 1140 starburst, three of which were located
in spectral regions well away from the OH sky lines, and clearly
resolved. In addition to strong [Fe{\sc ii}] emission (at
$\lambda_{\rm rest} = 1.644 \, \mu$m), we also detect the hydrogen
lines Br 12--4 ($\lambda_{\rm rest} = 1.641 \, \mu$m), Br 13--4
($\lambda_{\rm rest} = 1.611 \, \mu$m; although significantly affected
by a nearby strong sky line) and Br 14--4 ($\lambda_{\rm rest} = 1.588
\, \mu$m). We also detect a weak absorption feature at the wavelength
range expected for the CO(6,3) band head, in the most active starburst
regions (at $\lambda \simeq 1.627 \, \mu$m, after correcting for the
galaxy's redshift; see de Grijs et al. 2004). Its signal-to-noise
(S/N) ratio was not sufficient for reliable absorption-line studies,
however.

\begin{figure}
\begin{center}
\psfig{figure=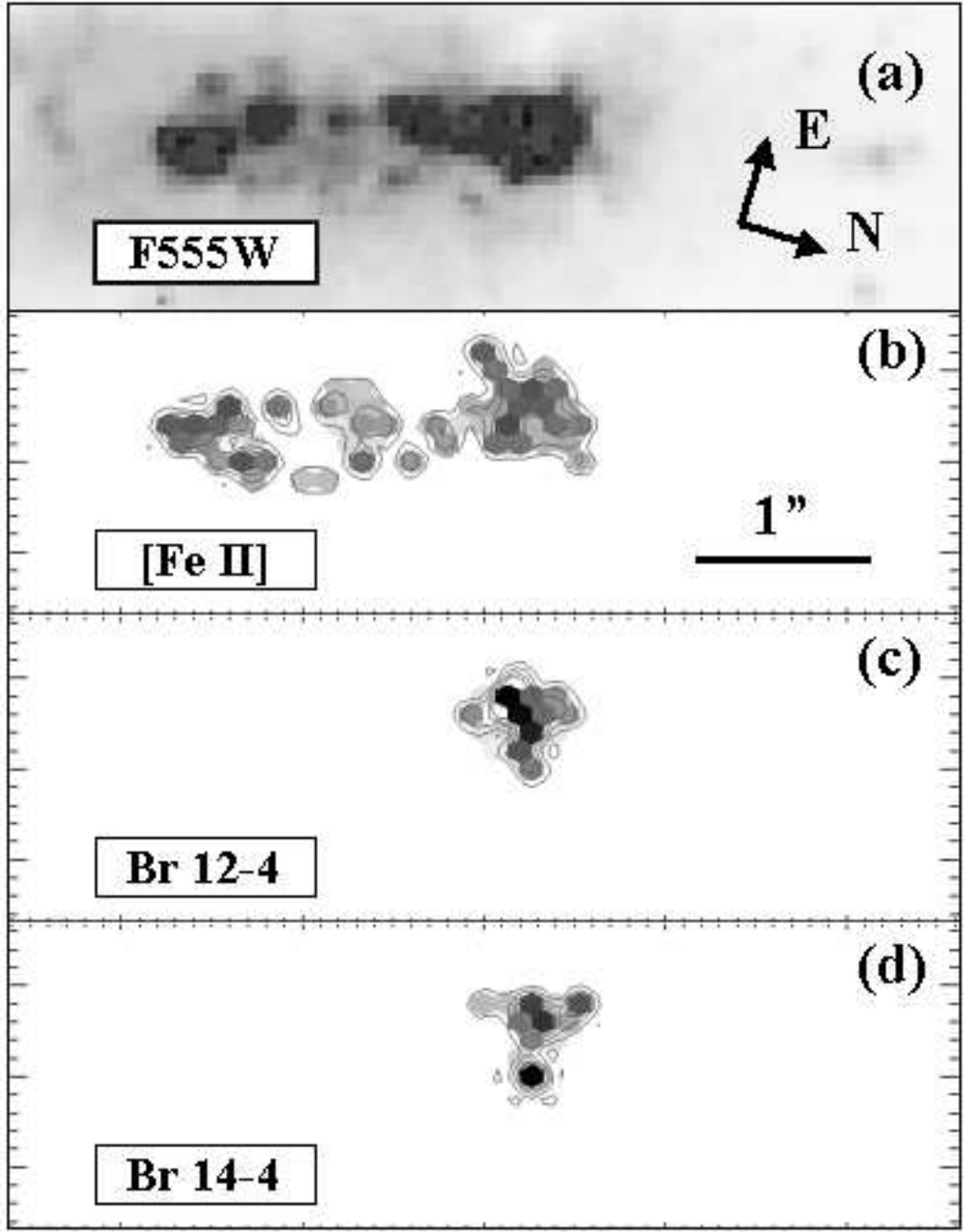}
\end{center}
\caption{\label{emission.fig}Maps of (a) the {\sl Hubble Space
Telescope} $V$-band continuum flux distribution, (b) the [Fe{\sc ii}]
emission, (c) the Br 12--4, and (d) the Br 14--4 emission-line
morphologies of NGC 1140 (from de Grijs et al. 2004). The CIRPASS data
show both the flux levels in the individual fibre lenses (above $\sim
3 \sigma$, where $\sigma$ is the noise in the continuum level of the
spectra near the respective emission lines), as well as smoothed
contours of the same data.}
\end{figure}

In Fig. \ref{emission.fig} we show the flux distribution of the
[Fe{\sc ii}], Br 12--4 and Br 14--4 emission lines across NGC 1140,
compared to the {\sl Hubble Space Telescope} $V$-band morphology.
While [Fe{\sc ii}] emission is found throughout the galaxy, the
emission from all of the higher-order Brackett lines (including Br
12--4, Br 13--4 and Br 14--4) is found predominantly associated with
the northern starburst region, NGC 1140-N. Although the observing
conditions under which the CIRPASS data were obtained were
non-photometric, the fact that we observe all wavelengths
simultaneously over the entire two-dimensional field of view enables
us to say with confidence that this different spatial distribution is
real. This is one of the major advantages of using IFUs.

The [Fe{\sc ii}] line on the one hand, and the Brackett 12--4 and
14--4 lines on the other originate in physically distinct processes,
each governed by their unique time-scale(s). [Fe{\sc ii}]$\lambda 1.64
\mu$m line emission originates through two different mechanisms.
Strong, compact [Fe{\sc ii}] emission originates in partially ionised
zones or shock-excited gas produced by supernovae (SNe). More diffuse
[Fe{\sc ii}] emission seems to be associated with galactic
superwinds. In starburst galaxies, compact [Fe{\sc ii}] emission is
thought to arise from discrete SN remnants (e.g., Oliva, Moorwood \&
Danziger 1989, Lumsden \& Puxley 1995, Vanzi \& Rieke 1997, Morel,
Doyon \& St-Louis 2002, Alonso-Herrero et al. 2003). Its excitation
requires both the destruction of dust grains (which contain a large
fraction of the interstellar iron) {\it and} large transition zones
between H{\sc ii} regions and regions of neutral hydrogen, in
association with a hard ionising source (so that the electron
temperature is sufficiently high to excite the forbidden [Fe{\sc ii}]
transitions; e.g., Oliva et al. 1989). Such spatial scales therefore
imply that thermal shock excitation is favoured (Oliva et al. 1989,
Vanzi \& Rieke 1997).

Brackett lines, on the other hand, are formed by recombination in
ionised gas associated with sources having a strong Lyman continuum,
such as H{\sc ii} regions created by young, massive stars. The strong
Brackett line emission will fade effectively after $\sim 8$ Myr, while
the [Fe{\sc ii}] emission remains observable significantly longer; the
[Fe{\sc ii}]/Brackett line ratio is therefore age dependent.

This difference in origin is quantitatively supported by our line
strength measurements. While both the composite Br 12--4 and Br 14--4
emission lines in NGC 1140-N are unresolved, the [Fe{\sc ii}] lines
are clearly resolved. We measure {\it intrinsic} FWHMs (i.e.,
corrected for the widths of unresolved lines, 3.8{\AA}) of 5.9{\AA}
and 4.2{\AA} in starburst regions NGC 1140-S (the southern starburst)
and N, respectively. This implies that the velocity dispersions are
$\sim 75$ and $\sim 110$ km s$^{-1}$ in NGC 1140-S and N,
respectively. This is consistent with the scenario that we are seeing
outflows in the [Fe{\sc ii}] emission, most likely driven by the
combination of SNe and stellar winds associated with the starburst.

The last SNe in a quenched starburst region occur at a time comparable
to the longest lifetime of a SN progenitor after the end of the
starburst activity, $\sim 35-55$ Myr (see de Grijs et al 2004 for a
discussion). For NGC 1140, we conclude therefore that the presence of
strong [Fe{\sc ii}] emission throughout the central galaxy implies
that {\it most} (but perhaps not all, allowing for the presence of
unresolved [Fe{\sc ii}] emission outside the currently active
starburst regions) of the recent star-formation activity was induced
some $35-55$ Myr ago.

Since the Br 12--4 and Br 14--4 emission, concentrated around NGC
1140-N is associated with recombination processes in younger H{\sc ii}
regions, this implies that the most recent starburst event has
transpired in this more confined area of the NGC 1140 disc. Thus,
based on the [Fe{\sc ii}] versus Brackett line emission, we conclude
that a galaxy-wide starburst was induced several $\times 10^7$ yr ago,
with more recent starburst activity concentrated around the northern
starburst region, NGC 1140-N.

\section{Higher resolution: Spatial vs. spectral resolution improvements}

Although our poor-quality CIRPASS observations provided us with
interesting insights into the galaxy's global recent star-formation
history, they did not allow us to probe the physically interesting
parsec scales, nor to resolve the individual YMCs (for which a seeing
$<0.7$ arcsec FWHM is required). At a spatial resolution of $\lesssim
0.5-0.7$ arcsec, we would be able to (just) separate the stellar and
gaseous components in the active starburst regions, and thus place the
current starburst in the context of the evolution of this starburst
galaxy as a whole.

Based on higher spatial resolution data, which are increasingly likely
to become available in the near future -- in particular in view of
worldwide efforts to provide high-quality adaptive optics-enabled
acquisition of scientific data -- I expect to be able to zoom in into
the galaxy's active starburst, and distinguish between the YMCs and
the intercluster medium, and thus follow cluster winds and other
outflows triggered by the ongoing starburst. The origin of the
ubiquitous outflows and feedback processes in active starburst
galaxies is as yet a key open question in this very active field, the
answers to which will most likely be provided by adaptive
optics-assisted IFS.

Secondly, the $R \sim 3500$ spectral resolution of CIRPASS proved to
be just sufficient to determine overall radial velocities to within
several tens of km s$^{-1}$ for entire starburst regions (de Grijs et
al. 2004), but {\it insufficient} to determine the individual YMC
velocities, nor those of the gaseous ISM. While the $R \sim 3500$
resolution would in principle have been sufficient for our YMC age
determinations, at higher spectral resolution (e.g., $R \gtrsim 6000$)
we expect to more easily resolve the stellar and cluster winds and
outflows, as well as the general kinematics of the underlying stellar
disk.

Finally, perhaps the simplest approach to improving the scientific
results in this contribution is to obtain higher-S/N observations, and
use these to redo the analysis. For instance, the CO band head in the
$H$ band is present, but in our CIRPASS data of insufficient quality
to measure its equivalent width robustly nor reliably, due to the poor
quality and low S/N ratio of the observations at hand. I point out,
however, that the CO band head features are significantly stronger in
both the $J$ and $K$ bands, thus suggesting that these might be more
appropriate wavelength ranges to target in studies of the associated
red supergiant content of starburst galaxies.

Studies of starburst galaxies in the local Universe, particularly
those involved in and/or triggered by galaxy interactions, provide key
diagnostics in terms of the resulting time-scales of the
star-formation activity and of the interaction history. Both are
essential ingredients in our quest to understand the details of galaxy
evolution as such. Examples from the local Universe, which we can
study in great detail, have therefore the potential to provide key
insights into galaxy evolution at higher redshifts, where galaxy
interactions and starburst events were much more common than at the
present, relatively quiescent epoch.

\section*{Acknowledgements}
I am very grateful to Linda Smith, Rob Sharp, Jay Gallagher, and the
CIRPASS team at the Institute of Astronomy in Cambridge, for their
input and assistance during this project. Without them, this work
could not have been completed successfully in a timely manner.

\end{document}